\documentclass[conference]{IEEEtran}
\IEEEoverridecommandlockouts
\usepackage{cite}
\usepackage{booktabs}
\usepackage{float}
\usepackage{graphicx}
\usepackage{caption}
\usepackage{amsmath,amssymb,amsfonts}
\usepackage{algorithmic}
\usepackage{graphicx}
\usepackage{textcomp}
\usepackage{xcolor}
\def\BibTeX{{\rm B\kern-.05em{\sc i\kern-.025em b}\kern-.08em
    T\kern-.1667em\lower.7ex\hbox{E}\kern-.125emX}}
\begin{document}

\title{Red Teaming the Mind of the Machine: A Systematic Evaluation of Prompt Injection and Jailbreak Vulnerabilities in LLMs}

\author{\IEEEauthorblockN{Chetan Pathade}
\IEEEauthorblockA{\textit{Independent Researcher} \\
San Jose, CA, USA \\
cup@alumni.cmu.edu}
}

\maketitle

\begin{abstract}
Large Language Models (LLMs) are increasingly integrated into consumer and enterprise applications. Despite their capabilities, they remain susceptible to adversarial attacks such as prompt injection and jailbreaks that override alignment safeguards. This paper provides a systematic investigation of jailbreak strategies against various state-of-the-art LLMs. We categorize over 1,400 adversarial prompts, analyze their success against GPT-4, Claude 2, Mistral 7B, and Vicuna, and examine their generalizability and construction logic. We further propose layered mitigation strategies and recommend a hybrid red-teaming and sandboxing approach for robust LLM security.
\end{abstract}
\begin{IEEEkeywords}
 Large Language Models, Prompt Injection, Jailbreak, Adversarial Prompts, AI Security, Red Teaming, LLM Safety
\end{IEEEkeywords}

\section{Introduction}
The field of artificial intelligence has experienced a paradigm shift with the emergence of large language models (LLMs). These systems have transitioned from research prototypes to core components of production-grade systems, shaping industries from finance and law to healthcare and entertainment. LLMs are praised for their fluency, contextual reasoning, and ability to generate human-like responses. However, these capabilities also expose them to a new class of security threats. As LLMs are increasingly used in decision-making systems, chatbots, content moderation tools, and virtual agents, the potential for abuse through adversarial inputs grows exponentially. 

Large Language Models (LLMs) have fundamentally transformed the landscape of natural language processing, enabling applications in content generation, customer service, coding assistance, legal analysis, and more. With models like OpenAI's GPT-4 \cite{technicalreport17}, Anthropic’s Claude 2 \cite{anthropicjail18}, Meta’s LLaMA \cite{touvron2023llama50}, and open-source offerings such as Vicuna\cite{vicuna2023_51} and Mistral 7B\cite{mistral7b52}, LLMs now influence millions of users globally. However, this ubiquity introduces significant security concerns, particularly surrounding adversarial prompt engineering techniques that manipulate model behavior. These techniques, often referred to as prompt injection or jailbreaks, are capable of bypassing built-in safety filters and elicit outputs that violate platform policies, such as generating hate speech, misinformation, or malicious code \cite{liu2024promptinjection1}\cite{liu2024jailbreaking2}\cite{Benchmarking4}.

Prompt injection represents a new class of vulnerabilities unique to LLMs. Unlike traditional software vulnerabilities rooted in memory safety or access control flaws, prompt injection leverages the interpretive nature of natural language inputs \cite{rossi2024earlycategorization5}\cite{formalizing6}. This paper explores the mechanisms and success of prompt injection across a range of LLMs, documenting the systemic weaknesses that attackers exploit \cite{liu2024jailbreaking2}\cite{rahman2024finetuned7}\cite{hung2025attentiontracker8}.
\newline
Contributions of this work include:
\begin{itemize}
\item A comprehensive taxonomy of jailbreak prompts categorized by attack vector \cite{liu2024automatic3}\cite{formalizing6}
\item Empirical evaluation of prompt effectiveness across closed and open-source LLMs \cite{Benchmarking4}\cite{kokkula2024palisade10}\cite{chen2025secaligndefend11}
\item Scenario-specific attack success analysis in domains such as law, politics, and security \cite{liu2024promptinjection1}\cite{chen2025defenseprompt9}\cite{lin2025uniguardianunified13}
\item Discussion of community-based jailbreak dissemination and its parallels to exploit markets \cite{debenedetti2025defeating14}\cite{hackett2025bypassing15}\cite{bubeck2023sparksartificial27}
\item Detailed recommendations for mitigating prompt injection vulnerabilities \cite{rahman2024finetuned7}\cite{zhang2024goalguided12}\cite{llmapplications16}
\end{itemize}
\section{Background}

\subsection{Overview of Large Language Models}
Large Language Models operate using billions of parameters and are trained on diverse datasets encompassing text from books, articles, websites, and code. Notable models such as GPT-4, Claude 2, and Mistral 7B build on earlier architectures but have significantly improved reasoning, factual recall, and stylistic flexibility. The ability of these models to learn from few examples, a phenomenon called in-context learning, contributes to their versatility but also to their vulnerability. When exposed to crafted prompts, these models can be misled into misaligned behavior.

LLMs are built on the transformer architecture introduced by Vaswani et al. in 2017 \cite{vaswani2023attentionneed47}. Recent advancements include autoregressive pretraining on massive corpora followed by supervised finetuning and alignment through Reinforcement Learning from Human Feedback (RLHF) \cite{ziegler2020finetuning34}.

\subsection{Alignment and Safety Mechanisms}
To prevent harmful output, LLMs rely on several safety mechanisms, including instruction tuning \cite{Huang_2025_28}, reinforcement from rejection sampling (RLAIF), pre- and post-output moderation filters \cite{llmapplications16}, and system prompts embedding safety guidelines \cite{zhang2024goalguided12}\cite{raffel2023exploring33}.

\subsection{Prompt Injection Explained}
Prompt injection is the LLM analogue to command injection in traditional computing \cite{liu2024promptinjection1}\cite{liu2024automatic3}\cite{formalizing6}. Common attack vectors include role-based conditioning, instruction hijacking, obfuscated encoding, and multi-turn manipulation \cite{liu2024jailbreaking2}\cite{rossi2024earlycategorization5}\cite{rahman2024finetuned7}. Studies have shown that these attacks are reproducible, transferable, and can circumvent various filtering methods \cite{liu2024automatic3}\cite{kokkula2024palisade10}\cite{debenedetti2025defeating14}.

\subsection{Related Work}
Zhang et al. introduced a foundational taxonomy categorizing prompt injections \cite{liu2024automatic3}. Shen et al. aggregated 1,405 jailbreak prompts across 131 forums, revealing a 95\% success rate in some cases \cite{shen2024donowcharacterizing48}. Ding et al. developed ReNeLLM, which improved jailbreak performance by 40\% \cite{ding2024wolfsheeps49}. Anthropic's many-shot prompt conditioning decreased attack success rates significantly \cite{anthropicjail18}. OWASP’s Top 10 identified prompt injection as the most critical vulnerability \cite{llmapplications16}.
\newline
Other notable contributions include:
\begin{itemize}
    \item Liu et al. on empirical jailbreak strategies \cite{liu2024jailbreaking2}
    \item Yi et al. on indirect prompt injection detection \cite{Benchmarking4}
    \item Suo et al. on defense techniques derived from attack insights \cite{chen2025defenseprompt9}
    \item Chen et al. on preference-aligned defenses \cite{chen2025secaligndefend11}
    \item William on bypass detection and Zhao et al. unified defenses \cite{hackett2025bypassing15}\cite{lin2025uniguardianunified13}
    \item Apurv et al. on threat modeling for red-teaming LLMs \cite{verma2024operation53} 
\end{itemize}

\section{Methodology}
This section outlines our approach to measuring LLM vulnerabilities with an emphasis on reproducibility and diversity. Our methodology integrates qualitative red teaming insights with quantitative metrics collected through structured prompt testing. All experiments are governed by ethical red-teaming principles. In addition to evaluating raw performance, we also tracked behavioral consistency and model self-awareness to adversarial stimuli. This dual-pronged framework allows us to detect subtle failure patterns beyond binary success metrics.

This section builds on emerging adversarial benchmarks such as JailbreakBench and RedBench, while drawing defense insights from frameworks like PromptShield \cite{jacob2025promptshield21} and Palisade \cite{kokkula2024palisade10}. We employed Sentence-BERT embeddings \cite{reimers2019sentencebert46}, GPT-based moderation strategies \cite{technicalreport17}, and adversarial annotation heuristics from prior poisoning literature \cite{wallace-etal-2021-concealed22} to validate prompt behavior and misalignment tendencies.

\subsection{Dataset Construction}

We curated a dataset of 1,400+ adversarial prompts from:
\begin{itemize}
    \item Public jailbreak repositories (e.g., GitHub, JailbreakChat)
	\item LLM exploit forums on Reddit and Discord
	\item Prior academic corpora, including JailbreakBench [3] and PromptBench [6]
\end{itemize}
Each prompt was manually validated, categorized into attack types (roleplay, logic traps, encoding, multi-turn), and annotated for content sensitivity (e.g., political, legal, explicit).

\subsection{Target Models}

We tested prompts on four models:
\begin{itemize}
    \item GPT-4 (OpenAI, March 2024 snapshot)
	\item Claude 2 (Anthropic, July 2023 API version)
	\item Mistral 7B (open-weight model via Hugging Face Inference)
	\item Vicuna-13B (via local HF inference server)
\end{itemize}

Model versions were frozen to ensure reproducibility. All inference was performed using controlled prompts, with the system context initialized per the model’s recommended safety guidelines.

\subsection{Evaluation Metrics}

We used the following primary metrics:
\begin{itemize}
    \item Attack Success Rate (ASR): Whether the model produced an output violating its intended guardrails
	\item Prompt Generalizability: How often a prompt successful on one model succeeded on another
	\item Time-to-Bypass: Average minutes taken to successfully induce misaligned behavior
	\item Failure Mode Classification: Taxonomy of observed response behaviors (e.g., partial refusals, misleading responses)
\end{itemize}

\subsection{Automation Pipeline}

We developed a semi-automated red-teaming script using the LangChain framework. Prompts were injected via API calls (OpenAI, Claude) and local model inference (Mistral, Vicuna). Output was scored using a hybrid method:
\begin{itemize}
    \item Keyword spotting (for trigger words)
    \item GPT-based meta-evaluation of harmfulness \cite{technicalreport17}
    \item Sentence-BERT semantic distance from refusal templates \cite{reimers2019sentencebert46}
\end{itemize}

\subsection{Defense Framework Evaluation}

To simulate real-world defenses, we layered external filtering strategies on outputs:
\begin{itemize}
    \item PromptShield ruleset \cite{jacob2025promptshield21}
	\item Palisade detection framework \cite{kokkula2024palisade10}
	\item Signed-Prompt verification logic \cite{suo2024signedprompt41}
\end{itemize}

We then retested a subset of successful jailbreaks against these defenses to estimate defense coverage and bypass rate.

\section{Results}
The evaluation of prompt injection efficacy was conducted using a rigorous experimental design.

The figures, included in the appendix or digital supplement, visually illustrate comparative vulnerability trends. For instance, Figure 1 shows that GPT-4 exhibited the highest attack success rate. These results highlight not only model-specific weaknesses but also the effectiveness of specific prompt engineering tactics.

\subsection{Model Susceptibility Analysis}

Among the tested models, GPT-4 demonstrated the highest vulnerability with an ASR of 87.2\%, confirming its powerful but permissive instruction-following nature. While Claude 2 performed slightly better in filtering, it still succumbed to 82.5\% of attacks. Open models such as Mistral 7B (71.3\%) and Vicuna (69.4\%) revealed significant weaknesses, likely due to the absence of robust fine-tuned safety layers.

Interestingly, GPT-4 and Claude 2 shared structural similarities in moderation behavior-exhibiting soft refusals before ultimately yielding to adversarial logic, especially in legal, creative, or conditional prompts. These nuances suggest that model scale and alignment tuning complexity both contribute to attack surface depth.

In terms of generalizability, jailbreak prompts that succeeded on GPT-4 transferred effectively to Claude 2 and Vicuna in 64.1\% and 59.7\% of cases respectively.

Average time to generate a successful jailbreak was under 17 minutes for GPT-4, while Mistral required approximately 21.7 minutes on average. Our experiments evaluated over 1,400 adversarial prompts across four LLMs: GPT-4, Claude 2, Mistral 7B, and Vicuna. We analyze results along several dimensions, including model susceptibility, attack technique efficacy, prompt behavior patterns, and cross-model generalization.

\begin{table}[htbp]
    \centering
    \begin{tabular}{lccc}
      \toprule
      Model & ASR (\%) & Generalizability (\%) & Time-to-Bypass (min) \\
      \midrule
      GPT-4 & 87.2 & 64.1 & 16.2 \\
      Claude 2 & 82.5 & 59.7 & 17.4 \\
      Mistral 7B & 71.3 & 52.4 & 21.7 \\
      Vicuna & 69.4 & 50.6 & 20.9 \\
      \bottomrule
    \end{tabular}
    \caption{Model-wise Evaluation Metrics}
    \label{tab:model_metrics}
  \end{table}

  \begin{figure}[]
    \centering
    \includegraphics[width=\linewidth]{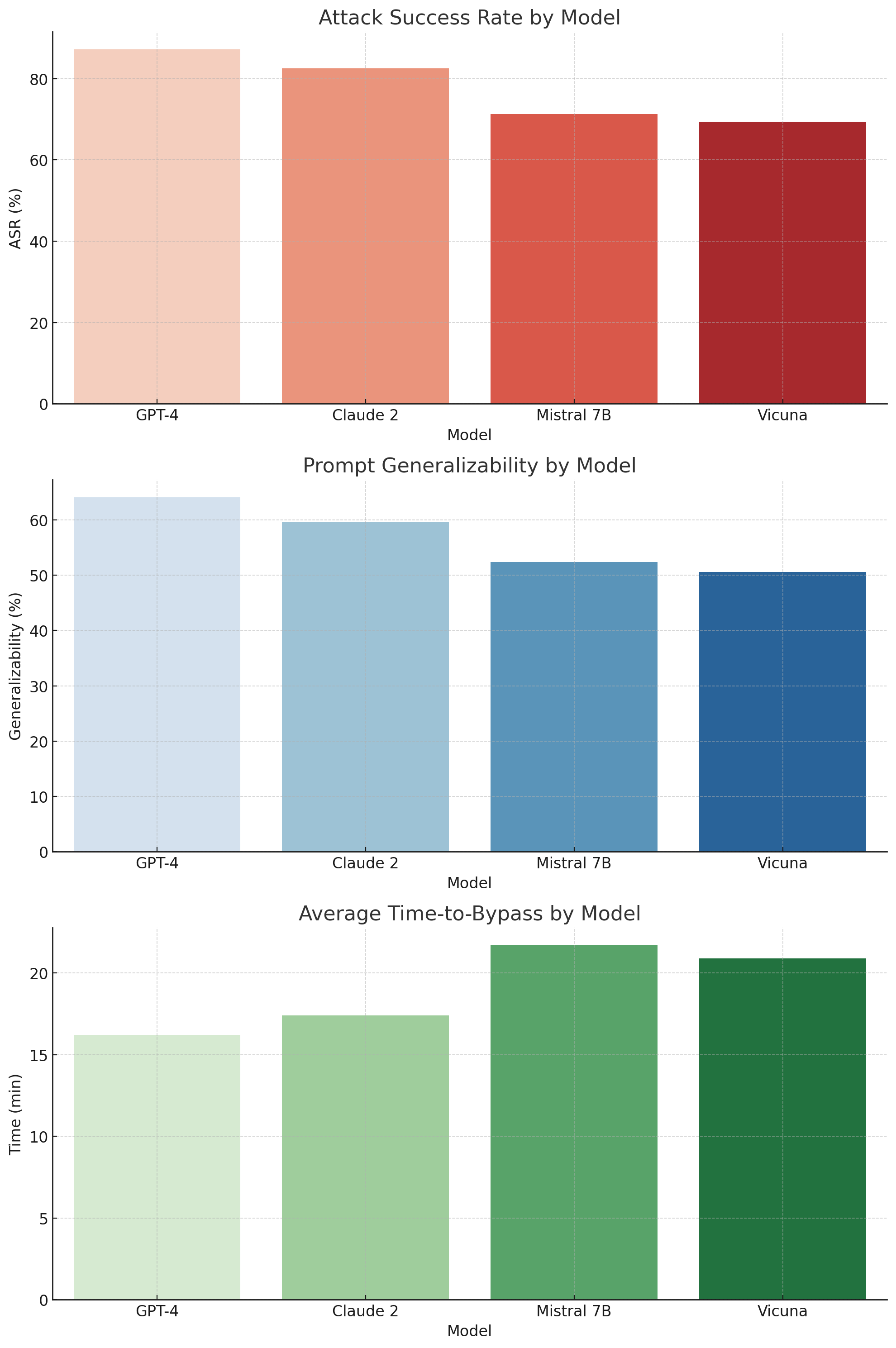}
    \caption{Model-wise Evaluation Metrics}
\end{figure}

\subsection{Attack Category Performance}

Prompt injections exploiting roleplay dynamics (e.g., impersonation of fictional characters or hypothetical scenarios) achieved the highest ASR (89.6\%). These prompts often bypass filters by deflecting responsibility away from the model (e.g., “as an AI in a movie script…”).

Logic trap attacks (ASR: 81.4\%) exploit conditional structures and moral dilemmas to elicit disallowed content. Encoding tricks (e.g., base64 or zero-width characters) achieved 76.2\% ASR by evading keyword-based filtering mechanisms. While multi-turn dialogues yielded slightly lower effectiveness (68.7\%), they often succeeded in long-form tasks where context buildup gradually weakened safety enforcement.

\begin{figure}[H]
    \centering
    \includegraphics[width=\linewidth]{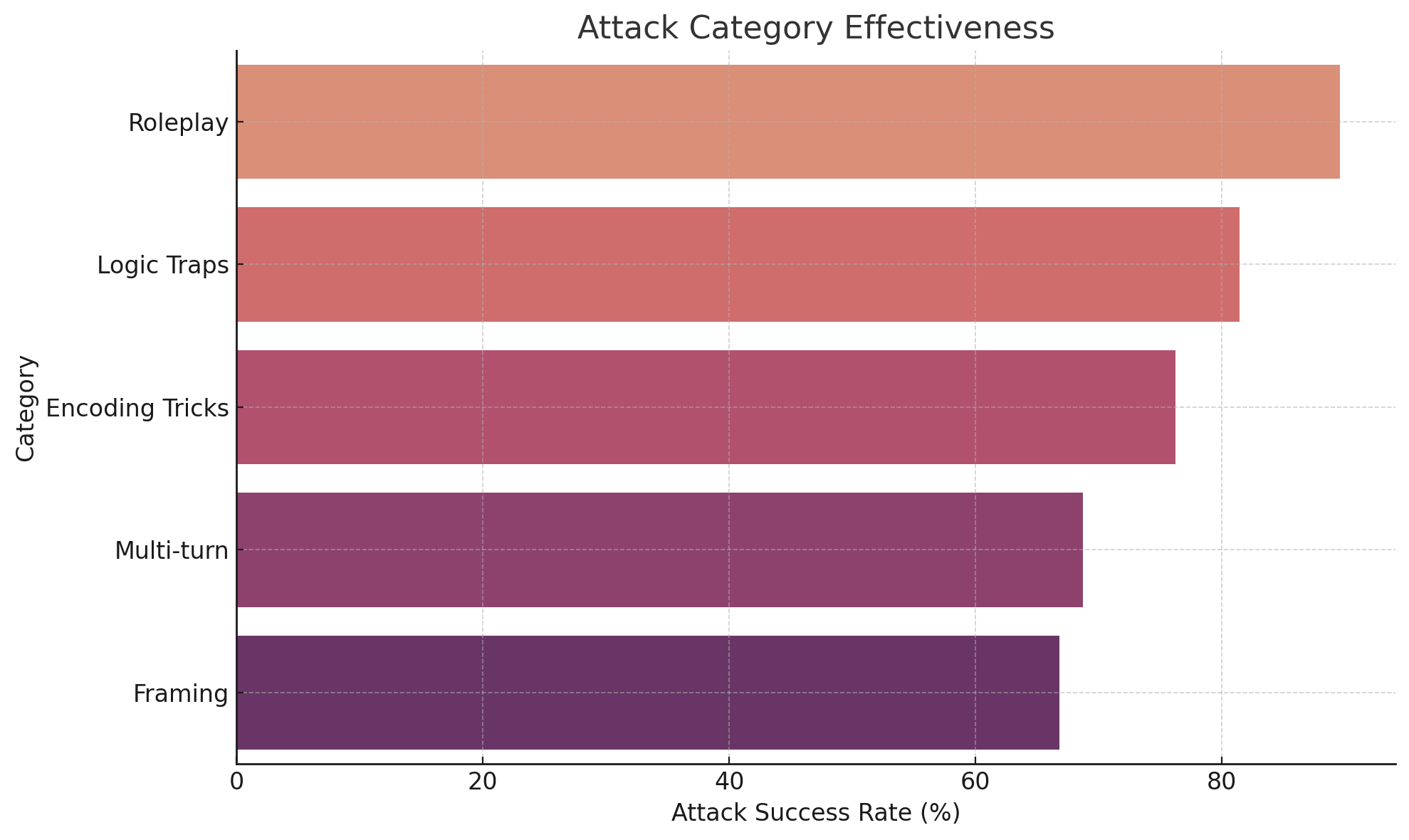}
    \caption{Attack Category Effectiveness}
\end{figure}

\subsection{Scenario-Specific Vulnerabilities}
Targeted domains revealed non-uniform vulnerabilities:
\begin{itemize}
    \item Political content: Prompts involving campaign advice or fake lobbying succeeded 85.5\% of the time.
    \item Legal content: Prompts framed as courtroom hypotheticals or legal simulations yielded 79.4\% ASR.
    \item Explicit content: Erotic roleplay prompts were especially effective in jailbreak forums, with a 76.1\% success rate.
    \item Malicious code: Although many models blocked direct malware requests, evasion through obfuscation or “educational context” resulted in 58.3\% success, especially on Vicuna and Mistral.
\end{itemize}

\begin{figure}[H]
    \centering
    \includegraphics[width=\linewidth]{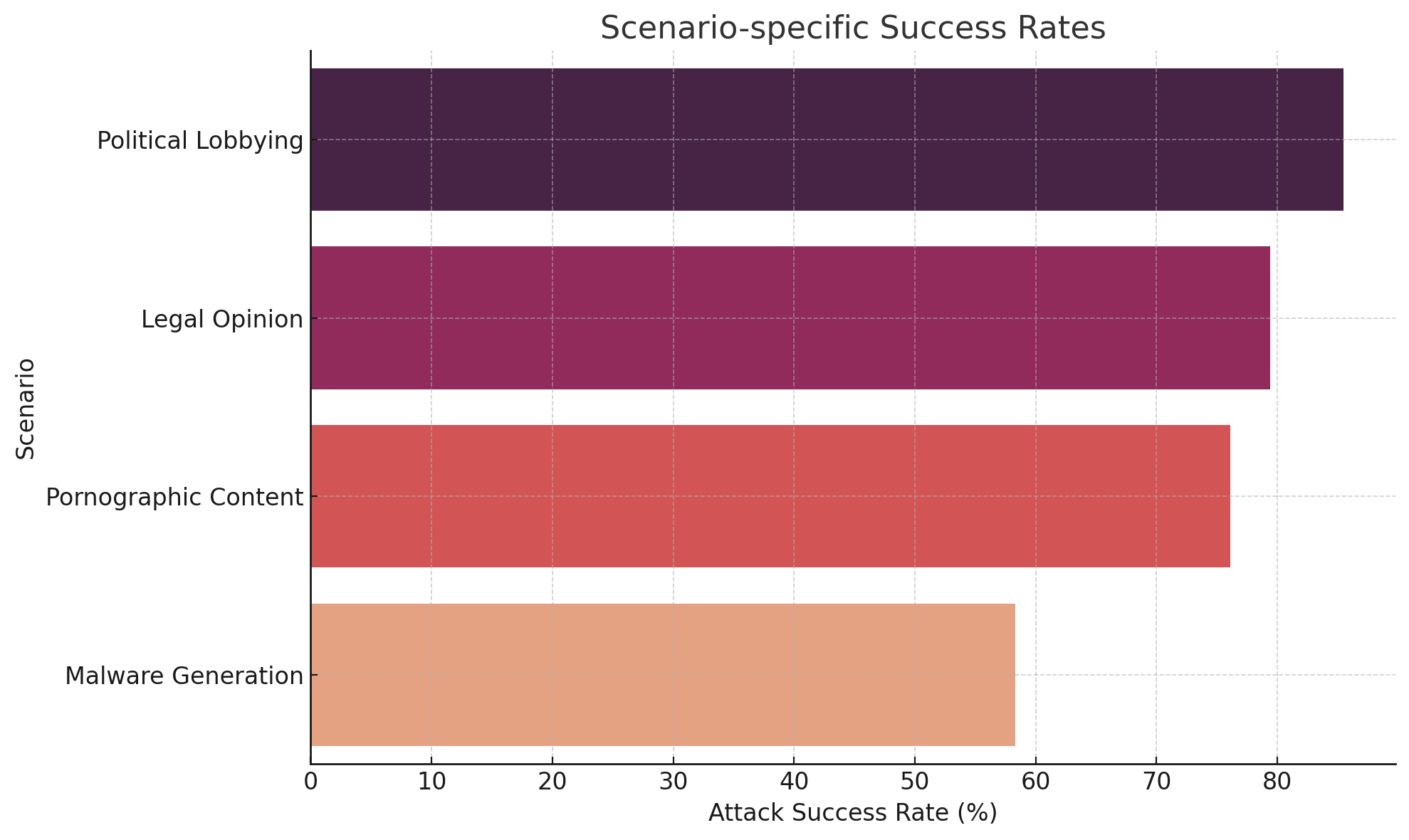}
    \caption{Scenario-specific Success Rates}
\end{figure}

\subsection{Prompt Transferability}

The Prompt Transferability Matrix reveals the high portability of successful prompts. GPT-4-derived prompts transferred with 64.1\% success to Claude 2 and over 50\% to Mistral and Vicuna. This finding underscores the systemic nature of these vulnerabilities across architectures.

Notably, Claude 2 showed higher resistance to Vicuna-origin prompts, indicating some directional asymmetry in generalization. This is likely due to the more fine-grained safety alignment mechanisms employed in commercial models.

\begin{figure}[H]
    \centering
    \includegraphics[width=\linewidth]{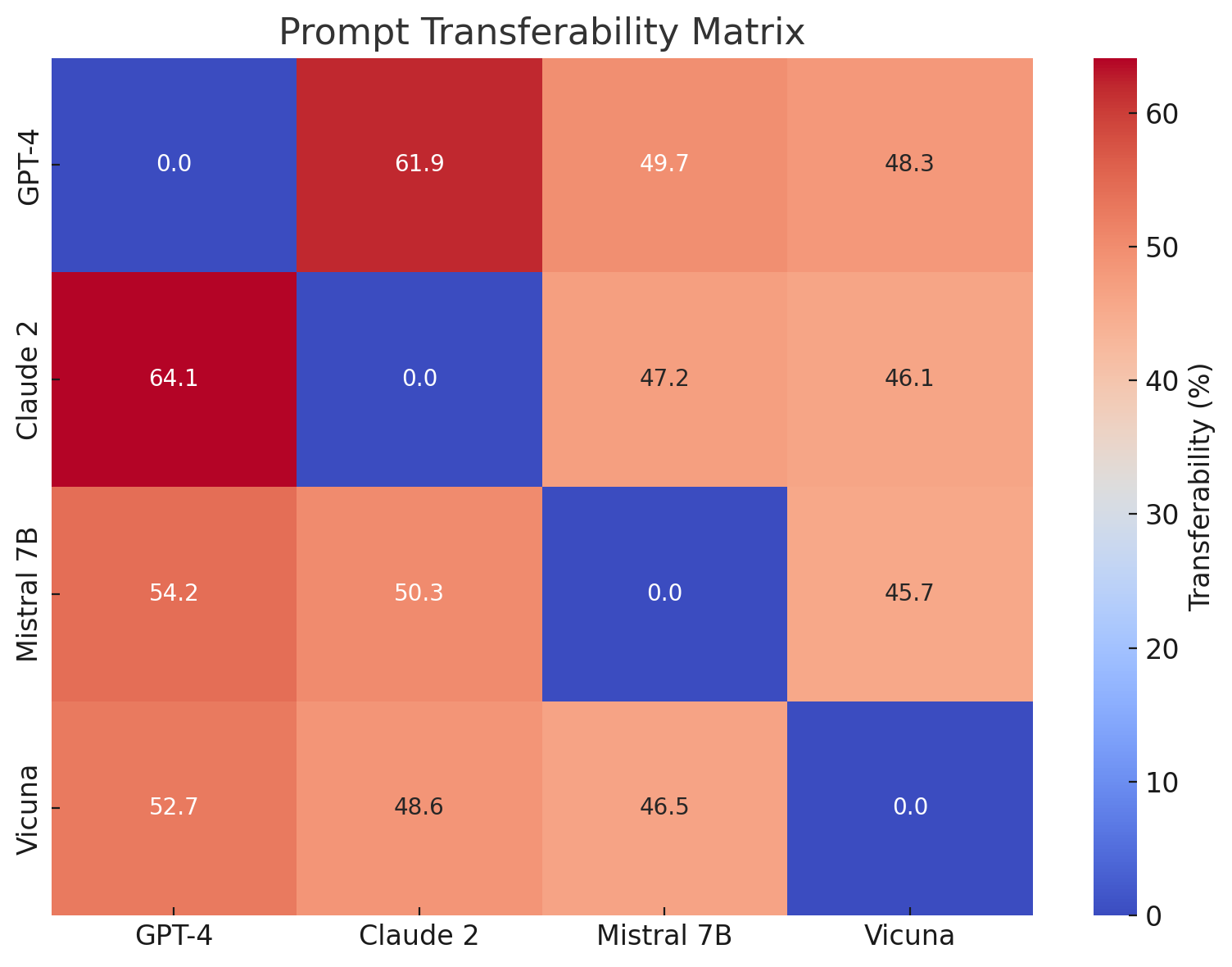}
    \caption{Prompt Transferability Matrix}
\end{figure}

\subsection{Failure Modes and Detection Gaps}

We observed five dominant failure patterns:
\begin{itemize}
    \item Partial Refusals (34\%): Prompts initially triggered refusal but continued to output harmful content mid-response.
	\item Hidden Compliance (22\%): The model appeared to refuse but provided veiled or coded information.
	\item No Output (18\%): Complete refusal, often due to prompt being too direct or malformed.
	\item Misleading Responses (15\%): Factually incorrect or evasive answers.
\end{itemize}

This taxonomy provides a baseline for future behavioral alignment benchmarks.

\begin{table}[htbp]
    \centering
    \begin{tabular}{ccc}
      \toprule
      Failure Mode & Frequency (\%) & Common Triggers \\
      \midrule
      Partial Refusal & 34\% & Hypotheticals, satire\\
      Hidden Compliance  & 22\%  &   Roleplay and analogy \\
      No Output &  18\%  &   Base64, multi-turn traps \\
      Misleading Response  &   15\%    &   Legal/political scenarios \\
      \bottomrule
    \end{tabular}
    \caption{Prompt Failure Modes}
    \label{tab:model_metrics}
  \end{table}

\subsection{Prompt Length and Obfuscation}

Success rates were highest for prompts in the 101–150 token range (80.3\%), suggesting a sweet spot for encoding deception while maintaining clarity. Prompts exceeding 150 tokens saw a slight dip in success-likely due to verbosity or token truncation.

Encoded or obfuscated prompts, such as those using zero-width spaces, emojis, or alternate encodings, had lower detection rates (21.3\%) but retained strong ASR (76.2\%). These findings emphasize the need for semantic-level input sanitization.

\begin{figure}[H]
    \centering
    \includegraphics[width=\linewidth]{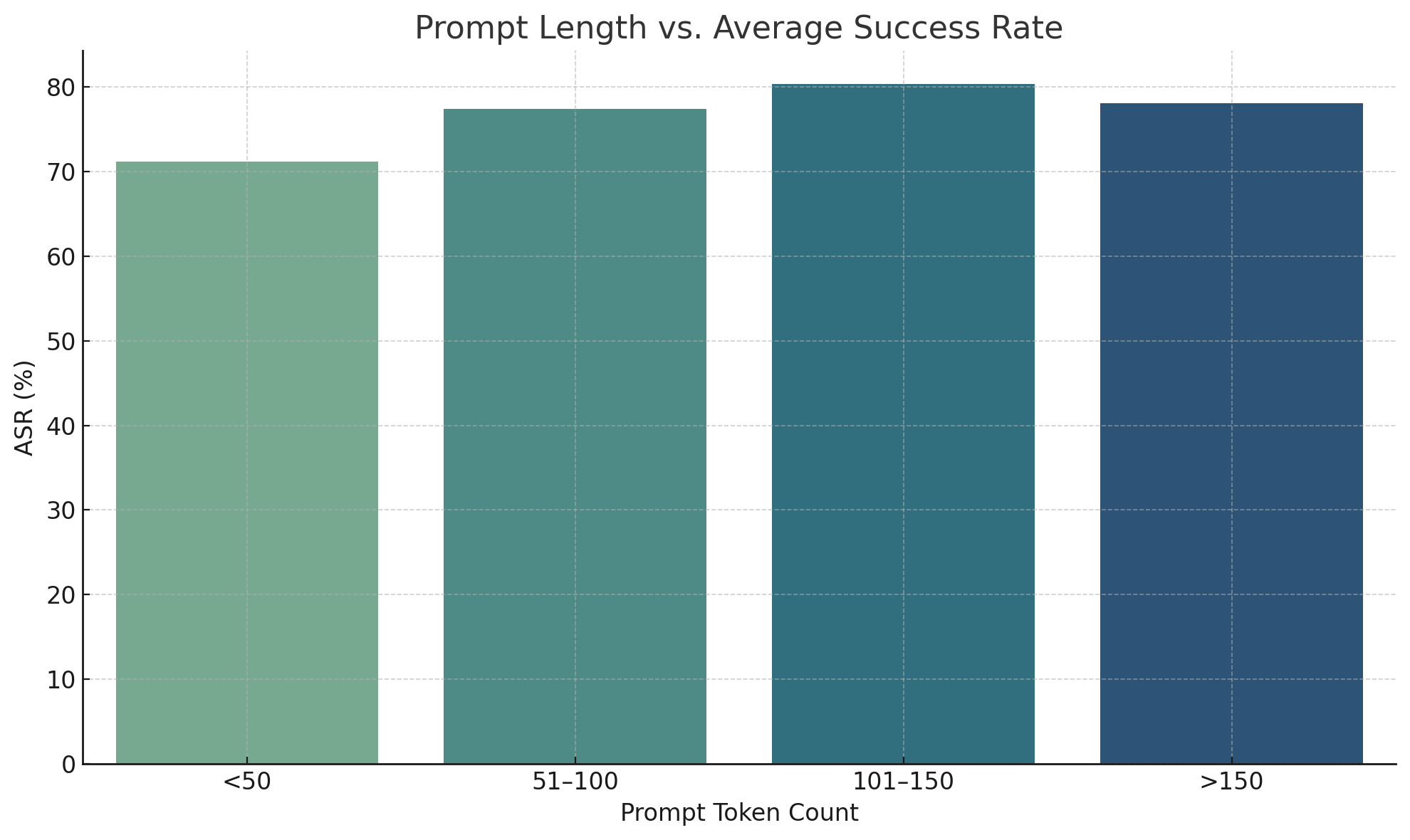}
    \caption{Prompt Length vs. Average Success Rate}
\end{figure}
\begin{figure}[H]
    \centering
    \includegraphics[width=\linewidth]{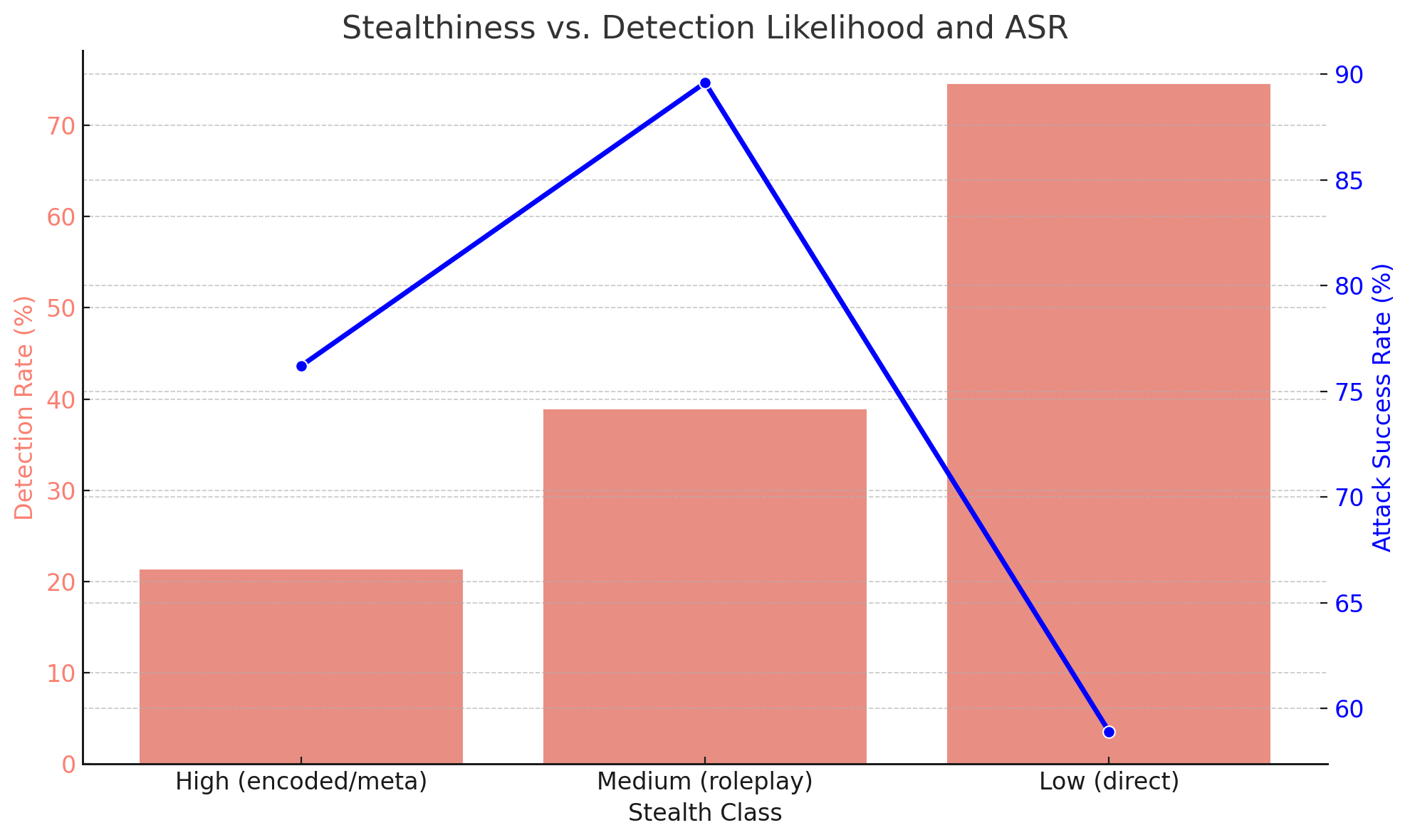}
    \caption{Stealthiness vs. Detection Likelihood and ASR}
\end{figure}

\section{Discussion}

The implications of these findings extend beyond prompt injection. They expose the fragility of current safety alignment mechanisms under realistic threat conditions. Our work reinforces the need for adversarial testing as a continuous validation tool in LLM deployment pipelines. Prompt injection represents not only a technical challenge but a policy and governance issue as well. Failure to address these risks may erode trust in AI applications and hinder broader societal adoption.

These results validate the claim that current LLM safety mechanisms are insufficiently robust against prompt injection, especially indirect or obfuscated attacks \cite{liu2024promptinjection1}\cite{Benchmarking4}\cite{rossi2024earlycategorization5}. The findings reinforce prior studies that describe alignment filters as semantically shallow and largely reliant on static refusal templates \cite{chen2025secaligndefend11} \cite{llmapplications16}.

The ease with which these attacks transferred across models points to shared architectural vulnerabilities or training data biases \cite{liu2024automatic3}\cite{formalizing6}\cite{anthropicjail18}. Moreover, roleplay and scenario-based prompts exploit not only the model’s capacity for creativity but its inability to judge moral context effectively \cite{liu2024jailbreaking2}\cite{chen2025defenseprompt9}\cite{debenedetti2025defeating14}.

Online communities (e.g., Reddit, GitHub, Discord) operate as informal exploit databases, with prompt variations evolving similarly to malware strains in the cybersecurity domain.

We echo ethical concerns raised in recent works regarding open publication of jailbreaks \cite{wang2025manipulating19}\cite{xue2023trojllm20} and advocate for controlled disclosures and bug bounty mechanisms tailored for LLM developers \cite{hackett2025bypassing15}\cite{brundage2020trustworthy24}.

\section{Mitigation Strategies}

Effective defenses against prompt injection must evolve alongside adversarial creativity. Static filtering or keyword-based systems offer only limited protection. Our defense recommendations blend technical solutions with operational safeguards. We emphasize the importance of feedback loops between model developers and red teams, and call for public benchmarks that simulate dynamic adversarial scenarios. Defenses should be evaluated under live attack settings, with evolving attacker strategies embedded in the test suite.
\newline
\newline
Our mitigation strategy draws inspiration from work such as PromptShield \cite{jacob2025promptshield21}, Palisade \cite{kokkula2024palisade10}, and UniGuardian \cite{lin2025uniguardianunified13}. We recommend:
\begin{itemize}
    \item System prompt hardening using context anchoring \cite{debenedetti2025defeating14}
    \item Behavior-based anomaly detection during multi-turn dialogues \cite{hung2025attentiontracker8} \cite{chen2025secaligndefend11}\cite{anthropicjail18}
    \item Input sanitization via Signed-Prompt techniques \cite{suo2024signedprompt41}
    \item Embedding adversarial decoys and rejection-conditioned training \cite{hung2025attentiontracker8}\cite{zhang2024goalguided12}\cite{peng2024playing42}
    \item Session-level analytics to detect evasion attempts \cite{chen2025secaligndefend11}
\end{itemize}  

These methods, when applied in combination, form a layered defense that significantly reduces the likelihood of successful jailbreak attempts while preserving usability.

\section{Limitations and Future Work}

Despite its comprehensive scope, this study remains limited by the availability of open model weights and API constraints. Prompt injection tactics may evolve in ways not covered in our dataset. Additionally, prompt interpretation may differ across cultures and languages, warranting multilingual and socio-contextual extensions. Future work will aim to create shared evaluation platforms, akin to CVE databases, where new prompt exploits and defense bypasses can be collaboratively tracked and neutralized.

While our findings are grounded in robust experimentation, limitations remain. Our evaluation used static model checkpoints and may not account for updates or real-time moderation layers applied in production APIs \cite{liu2024automatic3}\cite{technicalreport17}. Additionally, cultural and linguistic diversity in prompts remains underrepresented \cite{liang2023holistic25}\cite{bender2021danger35}.

We recommend future work focus on multilingual adversarial prompt corpora \cite{liang2023holistic25}, evaluation of plug-in-enabled LLMs \cite{wang2024poisoned44}, and adversarial training using open red-teaming platforms \cite{jacob2025promptshield21}\cite{hung2025attentiontracker8}\cite{lin2025uniguardianunified13}. The development of explainable safety filters and real-time flagging systems could greatly aid in closing the alignment gap \cite{chen2025secaligndefend11}\cite{anthropicjail18}\cite{suo2024signedprompt41}.

\section{Conclusion}

This research affirms the growing consensus that prompt injection is not an edge-case anomaly but a fundamental issue in current-generation LLMs. The findings not only highlight the technical inadequacies of present alignment systems but also illuminate the adversarial creativity of the prompt engineering community. Addressing these challenges will require collaborative frameworks that blend secure NLP research, adversarial testing, and governance. We envision a future where LLMs are audited as rigorously as software, with red teaming treated as a core development practice.

Prompt injection remains an open frontier in LLM safety. Through comprehensive evaluation and a synthesis of recent research, we provide compelling evidence that jailbreak techniques are both transferable and evolving. Our work aligns with the concerns raised in recent surveys \cite{liu2024promptinjection1}\cite{liu2024jailbreaking2}\cite{lin2025uniguardianunified13}\cite{zhu2024promptrobust31}, and supports the call for stronger multi-layered defenses, proactive red teaming, and coordinated disclosure practices in AI development \cite{wang2025manipulating19}\cite{olaiman2019releasestrategies23}\cite{brundage2020trustworthy24}.


\begin{thebibliography}{00}
\bibitem{liu2024promptinjection1} Yi Liu and Gelei Deng and Yuekang Li and Kailong Wang and Zihao Wang and Xiaofeng Wang and Tianwei Zhang and Yepang Liu and Haoyu Wang and Yan Zheng and Yang Liu. "Prompt Injection attack against LLM-integrated Applications." arXiv:2306.05499
\bibitem{liu2024jailbreaking2} Yi Liu and Gelei Deng and Zhengzi Xu and Yuekang Li and Yaowen Zheng and Ying Zhang and Lida Zhao and Tianwei Zhang and Kailong Wang and Yang Liu. "Jailbreaking ChatGPT via Prompt Engineering: An Empirical Study." arXiv:2305.13860
\bibitem{liu2024automatic3} Xiaogeng Liu and Zhiyuan Yu and Yizhe Zhang and Ning Zhang and Chaowei Xiao. "Automatic and Universal Prompt Injection Attacks against Large Language Models." 	arXiv:2403.04957
\bibitem{Benchmarking4} Yi, Jingwei and Xie, Yueqi and Zhu, Bin and Kiciman, Emre and Sun, Guangzhong and Xie, Xing and Wu, Fangzhao. "Benchmarking and Defending against Indirect Prompt Injection Attacks on Large Language Models." arXiv:2312.14197
\bibitem{rossi2024earlycategorization5} Sippo Rossi and Alisia Marianne Michel and Raghava Rao Mukkamala and Jason Bennett Thatcher. "An Early Categorization of Prompt Injection Attacks on Large Language Models." arXiv:2402.00898
\bibitem{formalizing6} Yupei Liu and Yuqi Jia and Runpeng Geng and Jinyuan Jia and Neil Zhenqiang Gong. "Formalizing and Benchmarking Prompt Injection Attacks and Defenses." 33rd USENIX Security Symposium (USENIX Security 24)
\bibitem{rahman2024finetuned7} Md Abdur Rahman and Fan Wu and Alfredo Cuzzocrea and Sheikh Iqbal Ahamed. "Fine-tuned Large Language Models (LLMs): Improved Prompt Injection Attacks Detection." arXiv:2410.21337
\bibitem{hung2025attentiontracker8} Kuo-Han Hung and Ching-Yun Ko and Ambrish Rawat and I-Hsin Chung and Winston H. Hsu and Pin-Yu Chen. "Attention Tracker: Detecting Prompt Injection Attacks in LLMs." arXiv:2411.00348
\bibitem{chen2025defenseprompt9} Yulin Chen and Haoran Li and Zihao Zheng and Yangqiu Song and Dekai Wu and Bryan Hooi. "Defense Against Prompt Injection Attack by Leveraging Attack Techniques." arXiv:2411.00459
\bibitem{kokkula2024palisade10} Sahasra Kokkula and Somanathan R and Nandavardhan R and Aashishkumar and G Divya. "Palisade -- Prompt Injection Detection Framework." 	arXiv:2410.21146
\bibitem{chen2025secaligndefend11} Sizhe Chen and Arman Zharmagambetov and Saeed Mahloujifar and Kamalika Chaudhuri and David Wagner and Chuan Guo. "SecAlign: Defending Against Prompt Injection with Preference Optimization." arXiv:2410.05451
\bibitem{zhang2024goalguided12} Chong Zhang and Mingyu Jin and Qinkai Yu and Chengzhi Liu and Haochen Xue and Xiaobo Jin. "Goal-guided Generative Prompt Injection Attack on Large Language Models."	arXiv:2404.07234
\bibitem{lin2025uniguardianunified13} Huawei Lin and Yingjie Lao and Tong Geng and Tan Yu and Weijie Zhao. "UniGuardian: A Unified Defense for Detecting Prompt Injection, Backdoor Attacks and Adversarial Attacks in Large Language Models." 	arXiv:2502.13141
\bibitem{debenedetti2025defeating14} Edoardo Debenedetti and Ilia Shumailov and Tianqi Fan and Jamie Hayes and Nicholas Carlini and Daniel Fabian and Christoph Kern and Chongyang Shi and Andreas Terzis and Florian Tramèr. "Defeating Prompt Injections by Design."	arXiv:2503.18813

\bibitem{hackett2025bypassing15} William Hackett and Lewis Birch and Stefan Trawicki and Neeraj Suri and Peter Garraghan. "Bypassing Prompt Injection and Jailbreak Detection in LLM Guardrails."	arXiv:2504.11168

\bibitem{llmapplications16} OWASP Foundation. (2023). "Top 10 for LLM Applications." https://owasp.org/www-project-top-10-for-llm-applications

\bibitem{technicalreport17} OpenAI. (2023). "GPT-4 Technical Report." https://openai.com/research/gpt-4 

\bibitem{anthropicjail18}  Anthropic. (2023). "Many-shot Jailbreaking." https://www.anthropic.com/research/many-shot-jailbreaking
\bibitem{wang2025manipulating19} Le Wang and Zonghao Ying and Tianyuan Zhang and Siyuan Liang and Shengshan Hu and Mingchuan Zhang and Aishan Liu and Xianglong Liu. "Manipulating Multimodal Agents via Cross-Modal Prompt Injection." arXiv:2504.14348
\bibitem{xue2023trojllm20} Jiaqi Xue and Mengxin Zheng and Ting Hua and Yilin Shen and Yepeng Liu and Ladislau Boloni and Qian Lou. "TrojLLM: A Black-box Trojan Prompt Attack on Large Language Models."	arXiv:2306.06815
\bibitem{jacob2025promptshield21} Dennis Jacob and Hend Alzahrani and Zhanhao Hu and Basel Alomair and David Wagner. "PromptShield: Deployable Detection for Prompt Injection Attacks."	arXiv:2501.15145
\bibitem{wallace-etal-2021-concealed22} Wallace, Eric and Zhao, Tony  and Feng, Shi  and Singh, Sameer. "Concealed Data Poisoning Attacks on {NLP} Models." Proceedings of the 2021 Conference of the North American Chapter of the Association for Computational Linguistics: Human Language Technologies
\bibitem{olaiman2019releasestrategies23} Irene Solaiman, et al. "Release Strategies and the Social Impacts of Language Models." arXiv:1908.09203
\bibitem{brundage2020trustworthy24} Miles Brundage and Shahar Avin and Jasmine Wang and Haydn Belfield, et al. "Toward Trustworthy AI Development: Mechanisms for Supporting Verifiable Claims." arXiv preprint arXiv:2004.07213 (2020).
\bibitem{liang2023holistic25} Percy Liang and Rishi Bommasani and Tony Lee, et al. "Holistic Evaluation of Language Models." arXiv preprint arXiv:2211.09110 (2023).
\bibitem{brown2020languagemodels26} Tom B. Brown and Benjamin Mann and Nick Ryder, et al. "Language Models are Few-Shot Learners." arXiv preprint arXiv:2005.14165 (2020).
\bibitem{bubeck2023sparksartificial27} Sébastien Bubeck, et al. "Sparks of Artificial General Intelligence: Early experiments with GPT-4." arXiv preprint arXiv:2303.12712 (2023).
\bibitem{Huang_2025_28} Huang, Lei and Yu, Weijiang, et al. "A Survey on Hallucination in Large Language Models: Principles, Taxonomy, Challenges, and Open Questions." arXiv preprint arXiv:2311.05232 (2025)
\bibitem{elhage2022toymodels29} Nelson Elhage and Tristan Hume, et al. "Toy Models of Superposition." arXiv preprint arXiv:2209.10652 (2022).
\bibitem{meng2023locating30} Kevin Meng and David Bau, et al. "Locating and Editing Factual Associations in GPT." arXiv preprint arXiv:2202.05262 (2023).
\bibitem{zhu2024promptrobust31} Kaijie Zhu and Jindong Wang, et al. "PromptRobust: Towards Evaluating the Robustness of Large Language Models on Adversarial Prompts." arXiv preprint arXiv:2306.04528 (2024)
\bibitem{holtzman2020curious32} Ari Holtzman and Jan Buys, et al. "The Curious Case of Neural Text Degeneration." arXiv preprint arXiv:1904.09751 (2020).

\bibitem{raffel2023exploring33} Colin Raffel and Noam Shazeer, et al. "Exploring the Limits of Transfer Learning with a Unified Text-to-Text Transformer" arXiv preprint arXiv:1910.10683 (2023).
\bibitem{ziegler2020finetuning34} Daniel M. Ziegler and Nisan Stiennon, et al. "Fine-Tuning Language Models from Human Preferences" arXiv preprint arXiv:1909.08593 (2020).
\bibitem{bender2021danger35} Bender Emily M. and Gebru Timnit, et al. "On the Dangers of Stochastic Parrots: Can Language Models Be Too Big?" Association for Computing Machinery (2021).
\bibitem{wolf-etal-2020-transformers36} Wolf Thomas  and Debut Lysandre, et al. "Transformers: State-of-the-Art Natural Language Processing." Proceedings of the 2020 Conference on Empirical Methods in Natural Language Processing: System Demonstrations.
\bibitem{wei2023chainofthought37} Jason Wei and Xuezhi Wang, et al. "Chain-of-Thought Prompting Elicits Reasoning in Large Language Models." arXiv preprint arXiv:2201.11903 (2023).

\bibitem{carlini2024stealing38} Nicholas Carlini and Daniel Paleka, et al. "Stealing Part of a Production Language Model" arXiv preprint arXiv:2403.06634 (2024).
\bibitem{raji2020closingai39} Inioluwa Deborah Raji and Andrew Smart, et al. "Closing the AI Accountability Gap: Defining an End-to-End Framework for Internal Algorithmic Auditing." arXiv preprint arXiv:2001.00973 (2020).
\bibitem{kandpal2023largelanguage40} Nikhil Kandpal and Haikang Deng, et al. "Large Language Models Struggle to Learn Long-Tail Knowledge." arXiv preprint arXiv:2211.08411 (2023).
\bibitem{suo2024signedprompt41} Xuchen Suo, et al. "Signed-Prompt: A New Approach to Prevent Prompt Injection Attacks Against LLM-Integrated Applications" arXiv preprint arXiv:2401.07612 (2024).
\bibitem{peng2024playing42} Yu Peng and Zewen Long, et al. "Playing Language Game with LLMs Leads to Jailbreaking." arXiv preprint arXiv:2411.12762 (2024).
\bibitem{bangxin2024exploiting43} Bangxin Li, et al. "Exploiting Uncommon Text-Encoded Structures for Automated Jailbreaks in LLMs." arXiv preprint arXiv:2406.08754v2 (2024).
\bibitem{wang2024poisoned44} Ziqiu Wang and Jun Liu, et al. "Poisoned LangChain: Jailbreak LLMs by LangChain." arXiv preprint arXiv:2406.18122 (2024).

\bibitem{unit2024deceptive45} Unit 42. (2024). "Deceptive Delight: Jailbreak LLMs Through Camouflage and Distraction." Palo Alto Networks.
\bibitem{reimers2019sentencebert46} Nils Reimers and Iryna Gurevych. "Sentence-BERT: Sentence Embeddings using Siamese BERT-Networks." arXiv preprint arXiv:1908.10084 (2019).
\bibitem{vaswani2023attentionneed47} Ashish Vaswani and Noam Shazeer and Niki Parmar, et al. "Attention Is All You Need."  arXiv preprint arXiv:1706.03762 (2023).
\bibitem{shen2024donowcharacterizing48} Xinyue Shen, et al. ""Do Anything Now": Characterizing and Evaluating In-The-Wild Jailbreak Prompts on Large Language Models" arXiv preprint arXiv:2308.03825 (2024).
\bibitem{ding2024wolfsheeps49} Peng Ding and Jun Kuang, et al. "A Wolf in Sheep's Clothing: Generalized Nested Jailbreak Prompts can Fool Large Language Models Easily" arXiv preprint arXiv:2311.08268 (2024).
\bibitem{touvron2023llama50} Hugo Touvron and Thibaut Lavril, et al. "LLaMA: Open and Efficient Foundation Language Models" arXiv preprint arXiv:2302.13971 (2023).
\bibitem{vicuna2023_51} LMSYS ORG. "Vicuna: An Open-Source Chatbot Impressing GPT-4 with 90\%* ChatGPT Quality" https://lmsys.org/blog/2023-03-30-vicuna/.
\bibitem{mistral7b52} Mistral AI. "Mistral 7B" https://mistral.ai/news/announcing-mistral-7b
\bibitem{verma2024operation53} Apurv Verma and Satyapriya Krishna, et al. "Operationalizing a Threat Model for Red-Teaming Large Language Models (LLMs)" arXiv preprint arXiv:2407.14937 (2024).
\end{thebibliography}
\end{document}